\documentclass[prb,
twocolumn, showpacs,
preprintnumbers,
amssymb, amsmath, floatfix]{revtex4}
\usepackage{graphicx}

\newcommand{\nn}{\ensuremath{\nonumber}}

\renewcommand{\vec}[1]{\ensuremath{\boldsymbol{\mathrm{#1}}}}

\newcommand{\pd}[2]{\ensuremath{\frac{\partial #1}{\partial #2}}}

\begin{document}
\title{c-axis optical conductivity from the Yang-Rice-Zhang model of the underdoped cuprates}%
\author{Phillip E. C. Ashby}
\email{ashbype@mcmaster.ca}
\affiliation{Department of Physics and Astronomy, McMaster University, Hamilton, Ontario, Canada L8S 4M1}

\author{J. P. Carbotte}
\email{carbotte@mcmaster.ca}
\affiliation{Department of Physics and Astronomy, McMaster University, Hamilton, Ontario, Canada L8S 4M1}
\affiliation{The Canadian Institute for Advanced Research, Toronto, Ontario, Canada M5G 1Z8}

\begin{abstract}
The c-axis optical response of the underdoped cuprates is qualitatively different from its in-plane counterpart.  The features of the pseudogap show themselves more prominently in the c-axis than in-plane.  We compute both the c-axis and in-plane optical conductivity using the Yang-Rice-Zhang model of the underdoped cuprates.  This model combined with coherent interlayer tunnelling is enough to explain the qualitative differences between the in-plane and c-axis data. We show how pseudogap features manifest themselves in the infrared and microwave conductivity within this model.
\end{abstract}

\maketitle

\section{Introduction}

The nature of the pseudogap phase in the underdoped cuprates is believed to be central to the understanding of high-$T_c$ superconductivity.\cite{Timusk:1999fk}  Many ideas have emerged to understand the origin of the pseudogap.  Examples of theories include the idea of preformed cooper pairs\cite{Norman:2007vn} or a competing order parameter, such as $d-$density wave order.\cite{Chakravarty:2001kx}  An alternative picture has its roots in Anderson's resonating valence bond (RVB) order.\cite{ANDERSON:1987fk,Zhang:1988uq}  Within the RVB framework the pseudogap can emerge naturally as one dopes a Mott insulating state with holes.\cite{Lee:2006kx}  The model of the underdoped cuprates by Yang, Rice, and Zhang (YRZ) is based on these ideas.\cite{Yang:2006ly}

Since the YRZ model was put forward, it has proved successful in describing many features of the underdoped cuprates that cannot be understood from conventional BCS theory.  The essential new feature is the presence of an additional energy scale, namely, the pseudogap.  In the YRZ model the pseudogap is responsible for reconstructing the antinodal portion of the Fermi surface into closed Luttinger pockets.  With this modification, the YRZ model has been able to qualitatively capture the physics of Raman Spectra, \cite{Valenzuela:2007kx, LeBlanc:2010vn} ARPES, \cite{Yang:2009ys,Yang:2011uq} specific heat, \cite{LeBlanc:2009zr} penetration depth, \cite{Carbotte:2010ly} as well as tunneling spectroscopy.\cite{Yang:2010ve}  More recently, we have applied the YRZ model to the c-axis transport properties where we showed that the YRZ model is able to explain the insulting-like c-axis behaviour while remaining metallic in-plane.\cite{Ashby:2013fk}  It has also been shown to account for the c-axis violation of the Ferrell-Glover-Tinkham sum rule.\cite{Carbotte:2012vn}  It is remarkable that such a simple modification is capable of capturing the physics of such a diverse range of topics. The recent ARPES observation of fully closed pockets in Bi2212\cite{Yang:2011uq} adds further support for the YRZ model. In fact, the results of Yang {\it et al.}\cite{Yang:2011uq} show that the size and shape of the Fermi pockets is in excellent agreement with the YRZ model.

The optical response of the c-axis is known to be dramatically different from the in-plane response, both in for infrared\cite{Homes:1993fk} as well as for microwave\cite{Hosseini:1998ys} frequencies. In this paper we examine the differences between in-plane and c-axis optical response.  We compute the optical conductivity using the YRZ formalism.  We show that the YRZ model is able to capture the qualitative behavior of the AC optical conductivity both in-plane, and along the c-axis.  We use the conductivity to extract information about the behaviour of the superfluid density, as well as the distribution of optical spectral weight.  Lastly, we use the low frequency portion of our data to extract the microwave conductivity and find good agreement with experimental findings.  In section \ref{sec:1} we introduce the formalism required to compute the optical conductivity within the YRZ model. We present our numerical results for the infrared optical conductivity in section \ref{sec:2}, and discuss optical sums and the microwave conductivity in section \ref{sec:3}.  We summarize and conclude in section \ref{sec:4}.

\section{Optical conductivity in the YRZ model of the underdoped cuprates} 
\label{sec:1}
The real part of the c-axis optical conductivity in the bubble approximation can be expressed in terms of the spectral density, $A(\vec{k},\omega)$, and the Gorkov anomalous spectral density, $B(\vec{k},\omega)$, through the Kubo formula:
\begin{widetext}
\begin{align}
\textrm{Re}\left[\sigma_{c}(\omega,T)\right] = -\frac{e^2d^2}{\omega}\sum_{\vec{k}}t_\perp^2(\vec{k})\int_{-\infty}^{\infty}\frac{d\omega'}{2\pi}\left[f(\omega'+\omega)-f(\omega')\right]\left[A(\vec{k},\omega')A(\vec{k},\omega'+\omega)+B(\vec{k},\omega')B(\vec{k},\omega'+\omega)\right].
\end{align}
\end{widetext}
Here $e$ is the electron charge, $f(\omega)$ is the Fermi distribution function, $t_\perp(\vec{k})$ is an interlayer hopping matrix element, and d is the interlayer distance. We will often make comparisons with the in plane conductivity where $t^2d^2$ should be replaced by $v_{k_x}^2$, the electron velocity.  The YRZ model provides the coherent part of the Greens function from which we can extract the required spectral densities. For a doping, $x$, the Greens function is given by
\begin{align}
G(\vec{k},\omega) = \sum_{\alpha=\pm}\frac{g_t(x)W_{\vec{k}}^\alpha}{\omega-E_{\vec{k}}^\alpha-\Delta_{sc}^2/(\omega+E_{\vec{k}}^\alpha)}.
\end{align}
In the above $g_t(x)$ is a Gutzwiller renormalization factor and is given by $g_t(x) = 2x/(1+x)$.  The two energy branches and weights are given by
\begin{align}
E^{\pm}_{\vec{k}} = \frac{1}{2}(\xi_{\vec{k}} - \xi_{\vec{k}}^0)\pm E_{\vec{k}},
\end{align}
and
\begin{align}
\label{eq:weight}W_{\vec{k}}^\pm = \frac{1}{2}\left(1\pm\frac{\tilde{\xi_{\vec{k}}}}{E_{\vec{k}}}\right).
\end{align}
Here
\begin{gather}
\label{eqn:en}E_{\vec{k}} = \sqrt{\tilde{\xi_{\vec{k}}}^2+\Delta_{pg}^2},\\
\tilde{\xi_{\vec{k}}} = \frac{\xi_{\vec{k}}+\xi_{\vec{k}}^0}{2},\\
\xi_{\vec{k}}^0 = -2t(x)(\cos k_x+\cos k_y),
\end{gather}
and
\begin{align}
\nn\xi_{\vec{k}} = \xi_{\vec{k}}^0 &-4t'(x)\cos k_x\cos k_y\\
&- 2t''(x)(\cos2k_x + \cos2k_y) - \mu_p.
\end{align}
\begin{figure}
\centering
  \includegraphics[width=0.8\linewidth]{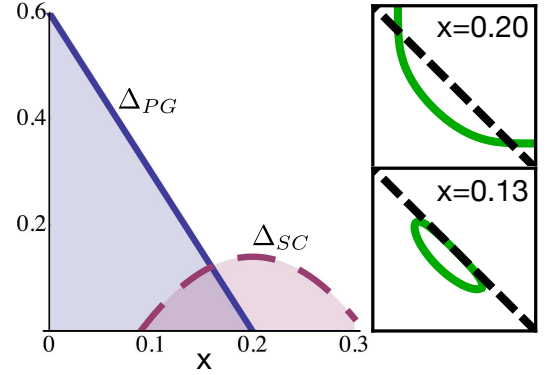}
    \caption{(Color online) The phase diagram and Fermi surface for the Yang-Rice-Zhang model. The quantum critical point at which the pseudogap emerges is set to $x = 0.2$ in this work, and corresponds to optimally doped superconductivity.  As the pseudogap grows with reduced doping the Fermi surface reconstructs from the large band structure Fermi surface at optimal doping into closed Luttinger pockets.}
  \label{fig:0}	
\end{figure}
$\mu_p$ is a chemical potential that is determined from the Luttinger sum rule. The hopping parameters are $t(x) = g_t(x)t_0+(3/8)g_s(x)J\chi$, $t'(x) = g_t(x)t_0'$, and $t''(x) = g_t(x)t_0''$,
where $g_s(x) = 4/(1+x)^2$ is another Gutzwiller renormalization factor, $J = t_0/3$, $\chi = 0.338$, $t_0'=-0.3t_0$ and $t_0''=0.2t_0$.  

In Eq. (\ref{eqn:en}) $\Delta_{pg}$ is the pseudogap energy scale which is taken to have d-wave symmetry, along with the superconducting gap.  That is,
\begin{align}
\Delta_{sc} = \frac{\Delta_{sc}^0(x)}{2}(\cos k_x-\cos k_y),\\
\Delta_{pg} = \frac{\Delta_{pg}^0(x)}{2}(\cos k_x-\cos k_y).
\end{align}
The doping dependent magnitudes mimic a simplified version of the cuprate phase diagram (see left panel of Fig. \ref{fig:0}):
 \begin{gather}
\Delta_{sc}^0(x) = 0.14(1-82.6(x-0.2)^2)\\
\Delta_{pg}^0(x) = 0.6(1-x/0.2).
\end{gather}
This functional form places both optimal doping (the maximum of the superconducting dome, shown in dashed purple in Fig. \ref{fig:0}) and the vanishing of the pseudogap energy scale (shown by the solid blue line in Fig. \ref{fig:0}) at $x=0.2$, in accordance with the original YRZ paper\cite{Yang:2006ly}.   Unless otherwise specified, we use all parameters in the band structure for the YRZ model as they appear in the original publication.\cite{Yang:2006ly}  In principle, one could alter these parameters to obtain fits to experimental data, but in this work we only wish to show that YRZ captures the essential physics.  For the magnitude of the superconducting gap, we use the ratio $2\Delta^0_{sc}(x,T=0)/(k_BT_c) = 6$. We work in units where $\hbar=1$, and measure all of our energies in terms of $t_0$, the nearest neighbour hopping amplitude.

\begin{figure}
\centering
  \includegraphics[width=0.8\linewidth]{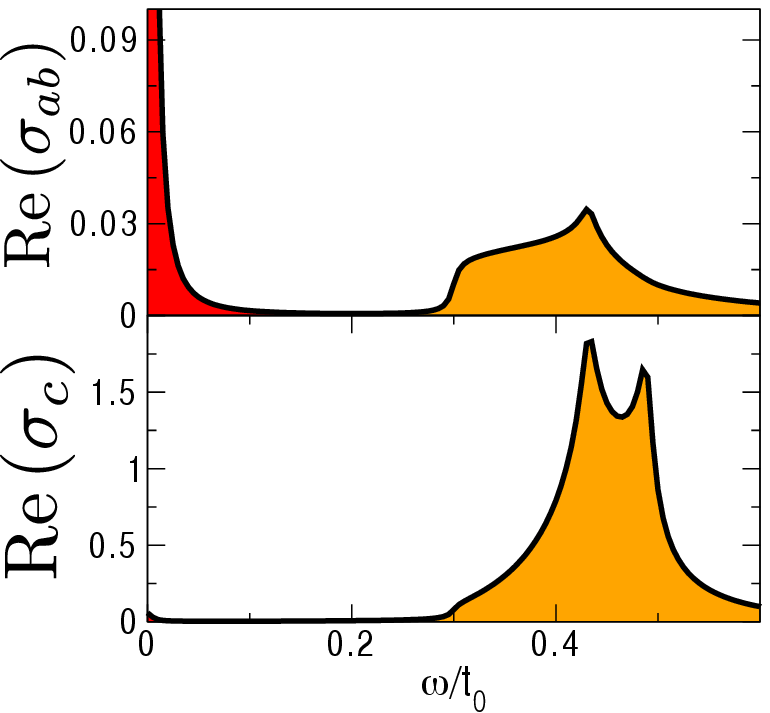}
  \includegraphics[width=0.8\linewidth]{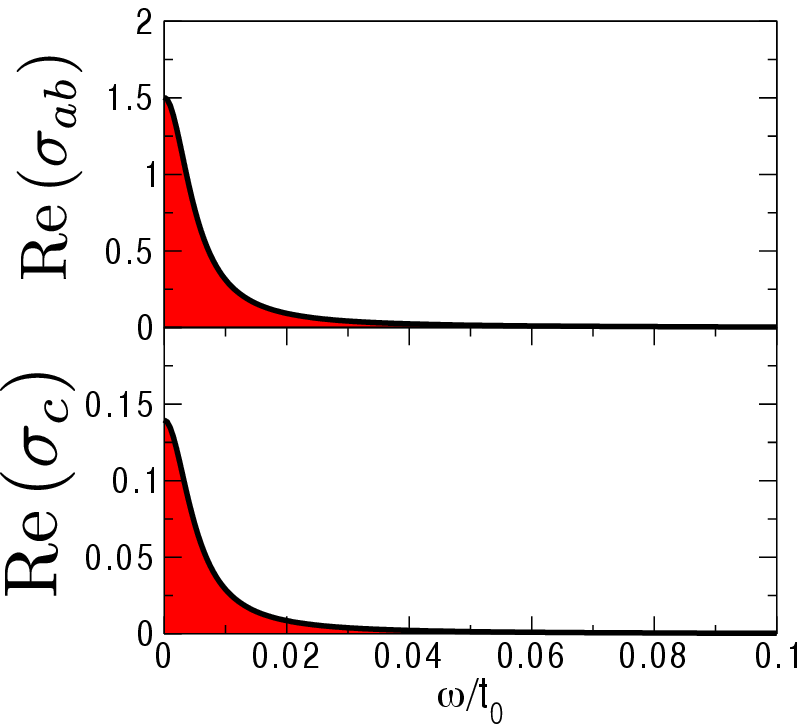}
    \caption{(Color online) (Top panel): The optical conductivity (Re($\sigma_{ab}$) measured in units of $e^2/d$, and Re($\sigma_c$) measured in units of $e^2dt_\perp^2$) at doping $x=0.13$ and $T=0.25T_c$. Both the ab-plane and c-axis contain a coherent Drude response (red) and an interband contribution from the presence of the pseudogap energy (orange). (Bottom panel): The optical conductivity at optimal doping ($x=0.20$) and $T=0.25T_c$. At optimal doping the pseudogap vanishes and there is only a Drude contribution to the conductivity.}
  \label{fig:1}	
\end{figure}

From the Greens function we obtain the spectral densities
\begin{align}
A(\vec{k},\omega) = 2\pi g_t(x)\sum_{\alpha= \pm}W_{\vec{k}}^\alpha&\left[(u^\alpha)^2\delta(\omega-E_s^\alpha)\right.\\
\nn&\left.+(v^\alpha)^2\delta(\omega+E_s^\alpha)\right],
\end{align}
and,
\begin{align}
B(\vec{k},\omega) = 2\pi g_t(x)\sum_{\alpha= \pm}W_{\vec{k}}^\alpha(u^\alpha v^\alpha)\left[\delta(\omega-E_s^\alpha)
-\delta(\omega+E_s^\alpha)\right].
\end{align}
The Bogoliubov quasiparticle energies, $E_s^\alpha$, and amplitudes, $u^\alpha$ and $v^\alpha$ are given by
\begin{align}
E_s^\alpha = \sqrt{E_{\vec{k}}^{\pm2}+\Delta_{sc}^2},\\
u^\alpha = \sqrt{\frac{1}{2}\left(1+\frac{E_{\vec{k}}^\alpha}{E_s^{\alpha}}\right)},
\end{align}
and
\begin{align}
v^\alpha = \sqrt{\frac{1}{2}\left(1-\frac{E_{\vec{k}}^\alpha}{E_s^{\alpha}}\right)}.
\end{align}

The right hand frame of Fig. \ref{fig:0} shows the normal state ($\Delta^0_{sc}(x) =0$) Fermi surface reconstruction brought about by the emergence of the pseudogap. Only the upper right quadrant of the Brillouin Zone (BZ) is shown.  The dashed black line indicates the antiferromagnetic BZ (AFBZ) boundary.   The solid green curve in the upper frame is the large Fermi surface of Fermi liquid theory for a doping of $x=0.2$. The Fermi liquid contour crosses the AFBZ but is unaffected by it and is characteristic of a good metal. As the doping is reduced towards half filling the Mott insulating state is approached and the Fermi surface contours change radically. The lower frame is for a doping of $x=0.13$ where the Fermi surface has reconstructed into a Luttinger hole pocket.  The backside of the Luttinger pocket at $x=0.13$ closely follows the AFBZ boundary and has very small weight ($W_k$ in Eqn. (\ref{eq:weight})). On the other hand, the front side is heavily weighted and is very close to the underlying Fermi liquid surface ($\Delta_{pg}=0$) in the nodal direction.  As $x$ is reduced further the Luttinger pocket continues to shrink and only a small number of well defined quasiparticles remain in the nodal direction.   This fact is very important for much of the physics that we will describe in this paper.

These small hole-like pockets are a prediction of the YRZ model and are in excellent agreement with recent photoemission data.\cite{Yang:2011uq}  However, it is believed that some of the transport properties in the pseudogap phase are electron-like.  Experimentally, it is found that both the Hall and Seebeck coefficients are negative in the pseudogap phase.  This apparent contradiction is nicely overcome in a recent proposal by Harrison and Sebastian.\cite{Harrison:2012fs}  They show how a nodal hole-like Fermi arc can be reconstructed into electron-like pockets.  Their reconstruction mechanism is due to bilayer charge ordering. The wavevectors associated with the charge modulation are responsible for the reconstruction from the hole-like Fermi surface into the electron-like one.  Importantly, the reconstruction happens with the nodal piece of Fermi surface in their model.  The YRZ model naturally makes hole pockets with long lived quasiparticles along the nodal direction.  Additionally, YRZ predicts that these hole pockets are heavily weighted along the front side of the pocket, and so this reconstruction mechanism should apply in exactly the same way as they describe.  This extra reconstruction offers a nice explanation for why these transport coefficients are observed as negative in this part of the phase diagram.

Using the above spectral functions, the conductivity can be written as the sum of two terms, $\textrm{Re}[\sigma] = \textrm{Re}[\sigma_D]+\textrm{Re}[\sigma_{IB}]$.  The first term, Re[$\sigma_D$], is peaked around $\omega=0$ and is a Drude-like response, while the second term, Re[$\sigma_{IB}$] arises from interband transitions between the different energy branches. In our calculations we take into account the effect of impurities by replacing the Dirac delta functions by Lorentzians of half-width $\Gamma$.  In the clean limit we find
\begin{widetext}
\begin{align}
\textrm{Re}[\sigma_D] = -2\pi e^2 d^2 g_t^2\sum_{\vec{k}}t^2_\perp(\vec{k})\delta(w)\left[W_{\vec{k}}^{+2}\pd{f(E_s^+)}{E_s^+}+W_{\vec{k}}^{-2}\pd{f(E_s^-)}{E_s^-}\right],
\end{align}
and
\begin{align}
\nn\textrm{Re}[\sigma_{IB}] =  2\pi e^2 d^2 g_t^2\sum_{\vec{k}}t^2_\perp(\vec{k})W_{\vec{k}}^+W_{\vec{k}}^-&\left[(u^-v^+-u^+v^-)^2\frac{1-f(E_s^+)-f(E_s^-)}{E_s^++E_s^-}[\delta(\omega-E_s^+-E_s^-)+\delta(\omega+E_s^++E_s^-)]\right.\\
&\left.-(u^+u^-+v^+v^-)^2\frac{f(E_s^+)-f(E_s^-)}{E_s^+-E_s^-}[\delta(\omega-E_s^++E_s^-)+\delta(\omega+E_s^+-E_s^-)]\right].
\end{align}
\end{widetext}
When using the clean limit formulas, one must instead replace the Dirac delta functions by Lorentzians of half-width $2\Gamma$.  

For the interlayer tunneling matrix element we follow Anderson \cite{Chakravarty:1993qf} and choose  $t_\perp(\vec{k})= t_\perp\left(\cos(k_x)-\cos(k_y)\right)^2$.  This choice reflects the geometric arrangement of the atoms between adjacent CuO$_2$ planes.  For the ab-plane the velocity is simply $v_{k_x} = d\xi/dk_x$.    The last free parameter we have is the scattering rate which broadens the Dirac delta functions.  We used 
\begin{align}
\label{eq:scatter}\Gamma = \begin{cases}
0.001+0.1\left(\frac{T}{T_c}\right)^3 & T \le T_c,\\
0.051+0.05\frac{T}{T_c} & T>T_c.\end{cases}
\end{align}
A linear in $T$ quasiparticle scattering rate has been associated with the inelastic scattering in the high $T_c$ oxides.  It is taken to be a characteristic of their normal state.  The marginal Fermi liquid phenomenology\cite{Varma:1989ly} is based on this observation as well as the idea that the the dominant scattering processes involve spin and charge excitations of the electronic system itself.  On entering the superconducting state, the emergence of the superconducting gap reduces both the charge and spin susceptibility and leads to a reduction in scattering. This reduction in scattering is a hallmark of an electronic mechanism for the inelastic scattering and is often referred to as the collapse of the inelastic scattering rate.\cite{Nuss:1991ve,Schachinger:1997qf,Nicol:1991bh,Hosseini:1999dq,Marsiglio:2006cr}  This collapse of the scattering rate is the accepted explanation of the large peaks observed in the microwave\cite{Bonn:1994nx} and thermal conductivity\cite{Matsukawa:1996oq} of the cuprates well below $T_c$.  While the normal fluid density (which is resistive) is reduced with temperature, the inelastic scattering lifetime increases.   The increase ceases when the residual scattering becomes dominant and it is the further reduction in normal fluid density that drives the conductivity to zero as observed in the experiment.\cite{Bonn:1994nx}

\section{Numerical Results}
\label{sec:2}
After all of these choices we can evaluate the conductivity. In all of our plots Re[$\sigma_{ab}$] is in units of $e^2/d$ while Re[$\sigma_c$] is measured in units of $e^2d t_\perp^2$.  This choice of units does not limit us to a particular material.  Once one chooses values for $t_\perp$, $d$, and the band structure parameters (which determine the Fermi velocity) then one can compare our calculations to any cuprate superconductor. Note that in these units, the in-plane conductivity need not be greater than its c-axis counterpart, which it is for any realistic value of Fermi velocity and tunneling matrix element. Fig. \ref{fig:1} shows the result of a calculation at $T = 0.25T_c$ for optimal doping ($x=0.20$) as well as underdoped ($x=0.13$).  We chose $x=0.2$ to highlight the physics in the absence of a pesudogap.  By contrast, the $x=0.13$ case has sizable pseudogap and superconducting energy scales.  This choice allows us to see the effect of both energy scales in our data most easily. In the $x=0.20$ case (bottom two frames of Fig. \ref{fig:1}), there is only a coherent Drude-like response from the large Fermi surface in both the ab-plane and the c-axis.  The c-axis response is reduced as compared with the ab response by the out of plane matrix element $t_\perp(\vec{k})$ which gives less weight to the part of the Fermi surface in the nodal direction.  In the underdoped case (top two frames of Fig. \ref{fig:1}), there is still a Drude response from the remaining Fermi surface.  It is suppressed in both the ab-plane and c-axis due to both the shrinking size of Fermi surface and the Gutzwiller renormalization factors. There is also a piece due to interband transitions at higher energies.  This piece is the signature of the pseudogap energy scale.  It is the dominant feature in the c-axis data.

\begin{figure}
  \includegraphics[width=0.8\linewidth]{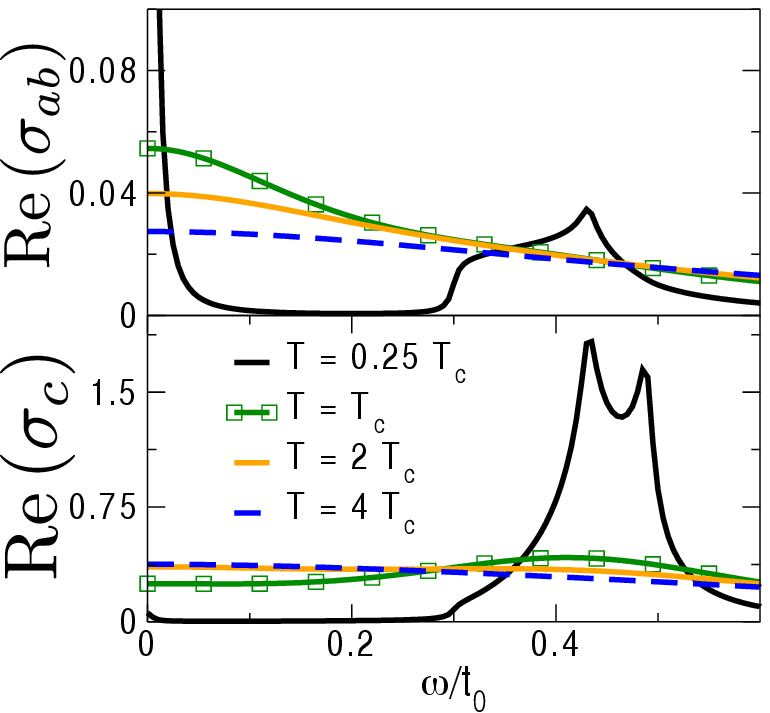}
    \includegraphics[width=0.8\linewidth]{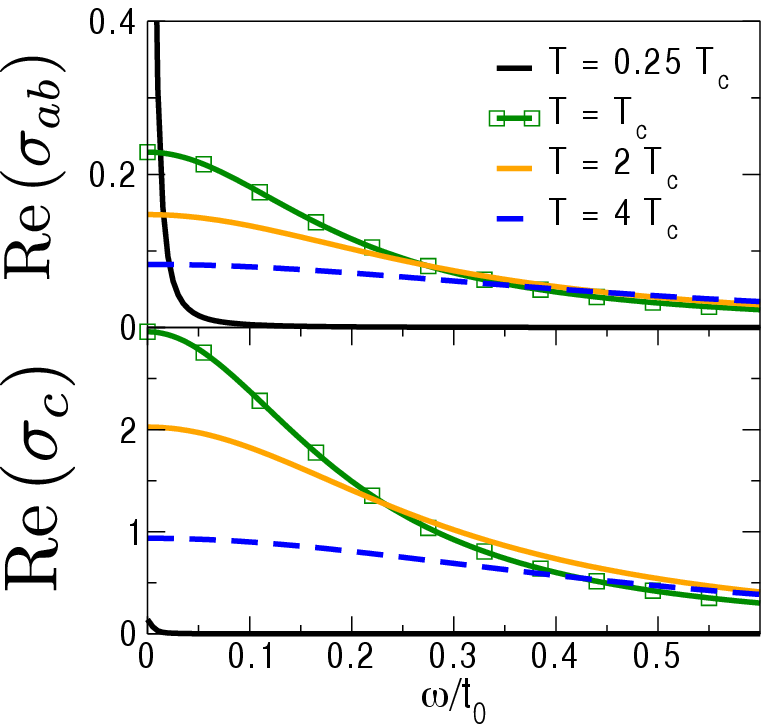}
    \includegraphics[width =0.8\linewidth]{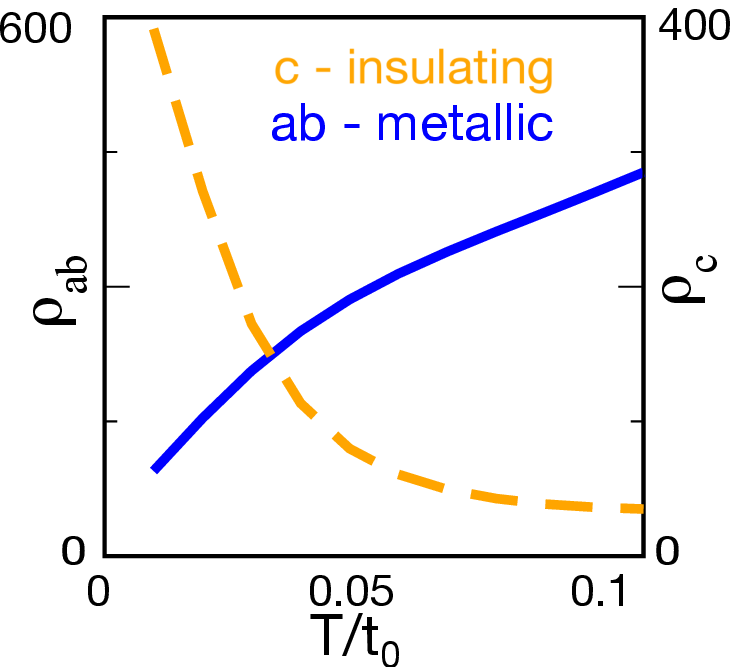}
    \caption{(Color online) (Top panel): The optical conductivity (Re($\sigma_{ab}$) measured in units of $e^2/d$, and Re($\sigma_c$) measured in units of $e^2dt_\perp^2$) at doping $x=0.13$ for temperatures $T = 0.25T_c, T_c, 2T_c,$ and $4T_c$.  The Drude peak becomes increasingly broadened with temperature. At low temperatures a feature of the pseudogap emerges beyond $\omega = 0.3t_0$.  This feature is a natural explanation for the broad peak observed at $400 \mathrm{cm}^{-1}$ in Homes {\it et al.}.\cite{Homes:1993fk} (Middle panel): The optical conductivity at doping $x=0.20$ for temperatures $T = 0.25T_c, T_c, 2T_c,$ and $4T_c$. Both in-plane and c-axis only have Drude contributions which become increasingly broadened with temperature. (Bottom panel):  The resistivity in the underdoped cuprates from the YRZ model as a function of temperature at $x=0.05$. The underdoped phase is metallic in-plane and resistive along the c-axis.  Taken with permission from Ashby and Carbotte.\cite{Ashby:2013fk}}
  \label{fig:1b}	
\end{figure}

In more realistic treatments of the inelastic scattering we expect this interband feature to broaden beyond what is shown in Fig. \ref{fig:1}.  In the cuprates the inelastic scattering rate is known to have a strong energy dependence\cite{Schachinger:2003kl} in addition to temperature variation.   In Eqn. (\ref{eq:scatter}) we have included only temperature dependence through an impurity model, i.e., there is no dependence on $\omega$.  This model for the inelastic scattering rate is perfectly adequate as far as DC properties are concerned. It is oversimplified when the photon energy falls in the infrared. In this energy range, $\Gamma(T,\omega)$ can be much larger than its $\omega = 0$ value, modeled in Eqn. (\ref{eq:scatter}).  Consequently, the interband optical transition peak in $\textrm{Re}[\sigma(t,\omega)]$ will be broadened. 

In Fig. \ref{fig:1b} the optical conductivity is shown as a function of frequency for four temperatures: as labeled in the figure, $T=4T_c$ dashed blue, $T=2T_c$ solid orange, $T=T_c$ in solid green with boxes and $T=0.25T_c$ in solid black.  The top two frames are for doping $x=0.13$ and the middle two at optimal doping ($x=0.2$).  For $\omega\lesssim 0.2t_0$ all curves are Drude like and order in temperature in the usual way, \emph{except} the c-axis curves in the underdoped case.  In this case, we see a different trend, the curves order in the opposite sense as temperature is reduced.  This behaviour is emphasized in the lowest frame where we show our results (for $x=0.05$) for the DC resistivity as a function of temperature.  We see that the c-axis displays insulating like behavior ($\rho_c$ increases as $T$ decreases), while the ab-plane response remains metallic ($\rho_{ab}$ decreases as $T$ decreases).  We stress that the YRZ model captures this behaviour though a coherent tunneling Hamiltonian.\cite{Levchenko:2010bh,Su:2006fk,Ashby:2013fk}

The top two frames of Fig. \ref{fig:1b} are to be compared with the experimental work of Homes {\it et al.} on the c-axis conductivity of YBCO.\cite{Homes:1993fk}  As in the experiments, the signatures of the pseudogap are more pronounced at low energies.  This is characterized by a flat region at low frequency which becomes increasingly suppressed at lower temperatures.  The reduction we calculate is not as dramatic as that observed in the experimental data. If one wished to obtain better fits to the experimental results it could be achieved by adjusting the scattering rate or the magnitude of the pesudogap energy scale.  Note that the low frequency c-axis conductivity behaves very differently from the in-plane conductivity as a function of temperature.  Perhaps the most striking feature in Fig. \ref{fig:1b} is the broad interband feature from $\omega \sim 0.3t_0$ to $\omega \sim 0.5t_0$ associated with the pseudogap.  This feature is a natural explanation for the broad peak observed in Homes {\it et al.}\cite{Homes:1993fk} at $400 \mathrm{cm}^{-1}$. Taking $t_0 = 125\mathrm{meV}$, this feature falls in the range $300-500\mathrm{cm}^{-1}$, in agreement with the experiments.  Previous explanations of this feature invoke the existence of an interlayer plasmon collective mode.\cite{Gruninger:2000uq}  Within YRZ this  peak is produced only from pseudogap physics with no need for this collective mode.

\section{Optical Sum and Microwave Conductivity}
\label{sec:3}
\subsection{Optical sums}

\label{sec:superflu}
\begin{figure}
\centering
  \includegraphics[width=0.8\linewidth]{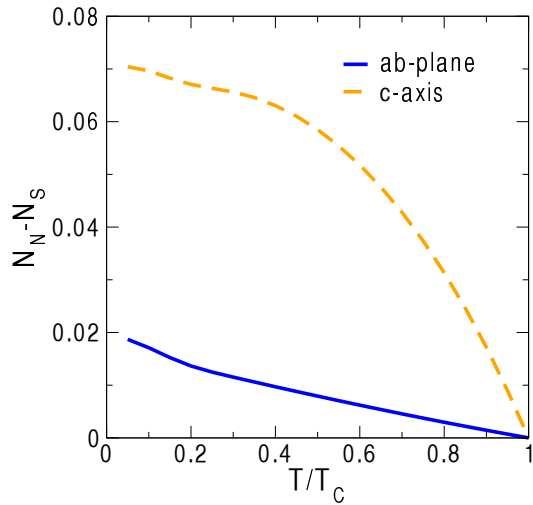}
    \caption{(Color online) Normal state optical sum minus superconducting state optical sum as a function of temperature for doping $x=0.13$. The behaviour of the superfluid density in the c direction is qualitatively different from the ab-plane.}
  \label{fig:4}	
\end{figure}

Sum rules and partial optical sums also provide useful information.   In a conventional superconductor, the suppression of $\mathrm{Re}[\sigma]$ for $T<T_c$ is connected to the appearance of a superconducting condensate.  The `missing' optical spectral weight appears in a delta function of the appropriate weight at $\omega = 0$.  This fact is usually presented as the Ferrel-Glover-Tinkham sum rule.  In terms of the superfluid stiffness, $\rho_s$:
\begin{align}
N_N-N_S = \rho_s,
\end{align}
where
\begin{align}
N_N = \int_{0^+}^\infty d\omega \mathrm{Re}[\sigma_n(\omega,T)]
\end{align}
is the normal state optical sum, and
\begin{align}
N_S = \int_{0^+}^\infty d\omega \mathrm{Re}[\sigma_s(\omega,T)]
\end{align}
is the superconducting optical sum. While this sum rule holds in-plane, it is known to be violated in the c-axis.\cite{Basov:1999zr} The YRZ model with the same interlayer tunneling matrix element that we use displays a violation of this sum rule.\cite{Carbotte:2012vn}  This is understood most simply in a limit of YRZ that reduces to a Fermi arc model.  In this model the Fermi surface is confined about the nodal direction; the remainder of the large Fermi surface from Fermi liquid theory is gapped out by the pseudogap.  Only the electrons on the arc contribute the usual amount to the optical sum rule, while those on the gapped out portion contribute less. In any case, $N_N - N_S$ is well defined and is shown in Fig. \ref{fig:4}.  The solid blue line is the ab-plane result, while the orange dashed line applies to the c-direction.  As a function of the reduced temperature, $t=T/T_c$ those curves behave in much the same way is is found for the superfluid density itself.\cite{Carbotte:2010ly,Carbotte:2013kx}  The superfluid density follows from the imaginary part of the conductivity as
\begin{align}
\frac{1}{\lambda^2(T)} = \lim_{\omega\rightarrow0}\frac{4\pi}{c^2}\omega \textrm{Im}[\sigma(T,\omega)],
\end{align}
with $c$ the velocity of light.  The temperature behaviours are in agreement with the superfluid density inferred from microwave experiments.\cite{Hosseini:1998ys}  The ab-plane superfluid density decreases linearly with temperature while the c-axis is flat at low temperatures.

\begin{figure}
\centering
  \includegraphics[width=0.8\linewidth]{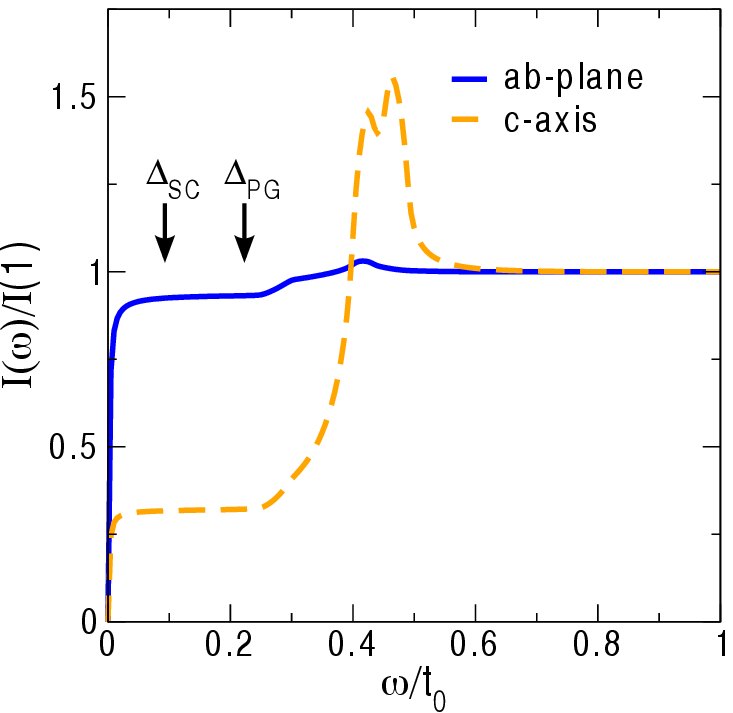}
    \includegraphics[width=0.8\linewidth]{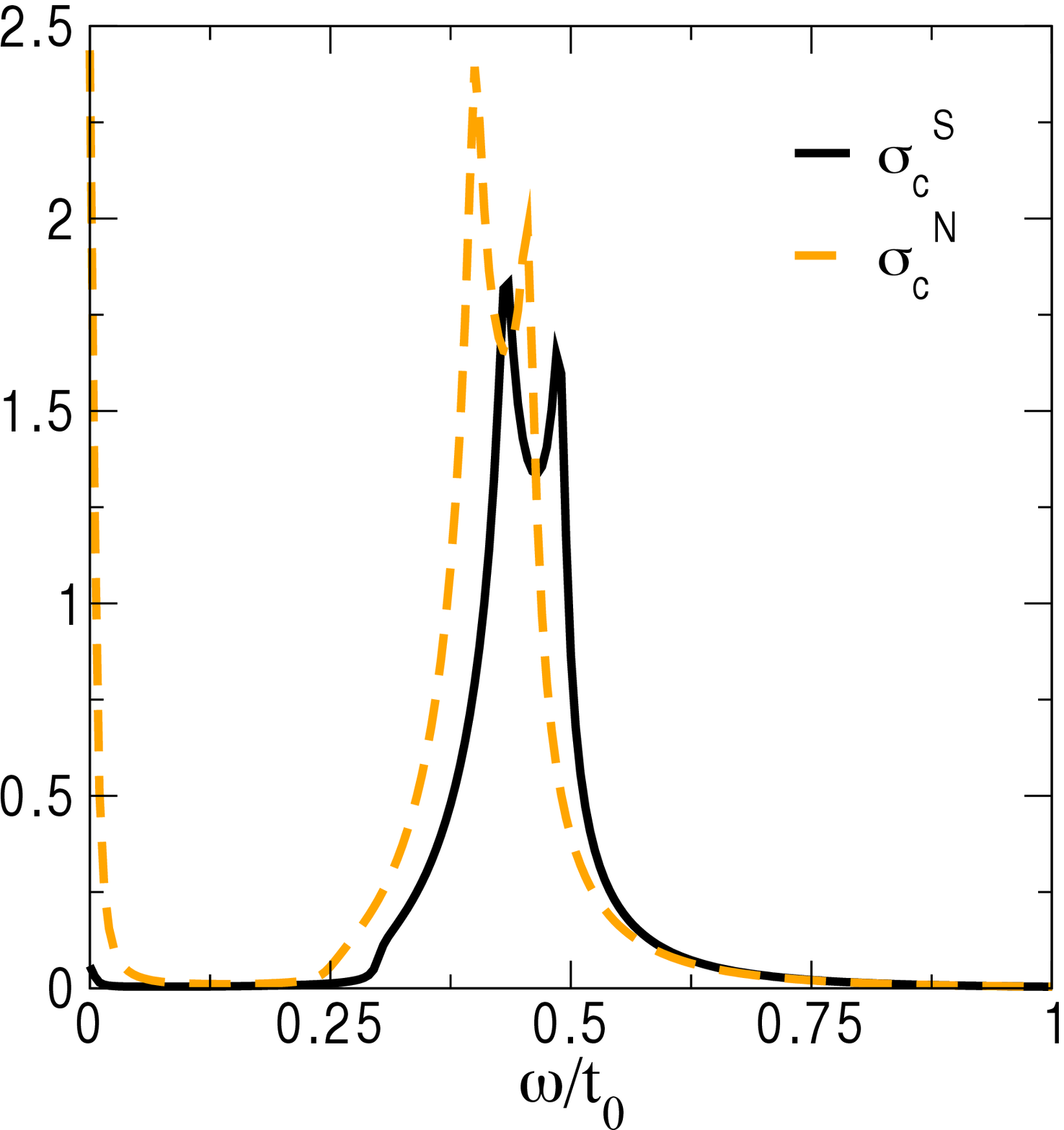}
    \caption{(Color online)  (Top Panel): Partial optical sum at $T = 0.25T_c$ for doping $x=0.13$ for the ab-plane and the c-axis.  The  gap scales are indicated on the Fig.. The pseudogap causes an extra suppression and redistributes the charge carriers available for condensation to high frequencies. (Bottom Panel): c-axis conductivities in the superconducting and normal states for $x=0.13$ at $T=0.25T_c$. The redistribution of spectral weight due from having superconductivity leads to a shift in the interband peak.}
  \label{fig:5}	
\end{figure}

A related quantity, the partial optical sum
\begin{align}
I(\omega) = \int_{0^+}^\omega d\omega'  \left(\mathrm{Re}[\sigma_n(\omega',T)] -  \mathrm{Re}[\sigma_s(\omega',T)]\right),
\end{align}
provides information about the distribution of spectral weight that goes into the superconducting condensate.  The partial optical sums for the YRZ model, normalized to their value at $\omega = 1$ are shown in the top panel of Fig. \ref{fig:5}.  The solid blue curve is for the ab-plane and the dashed orange curve is for the c-axis.  For a regular superconductor this curve would sharply rise towards 1, with the scattering rate, $\Gamma$, setting the energy scale of the rise.  In the YRZ model the presence of the pseudogap causes a redistribution of spectral weight from the Drude to the high frequency region.  This redistribution shows itself in the suppressed flat region at low frequencies accompanied by a pile up of spectral weight above the pseudogap energy scale.  This redistribution is much more pronounced for the c-axis.  Insight into these shifts in spectral weight can be gleaned from the lower frame where we show the real part of the conductivity as a function of frequency for $x=0.13$ at $T=0.25T_c$.  The dashed orange curve is the normal state ($\Delta_{pg}= 0$) and the black curve is the corresponding superconducting case.  We see that the opening of the superconducting gap shifts the interband transitions to higher energies. This shift accounts for the large peak seen in the orange curve in the top frame of Fig. \ref{fig:5} for $\omega\gtrsim 0.4t_0$.

\subsection{Microwave conductivity}
\label{sec:micro}
\begin{figure}
\centering
  \includegraphics[width=0.8\linewidth]{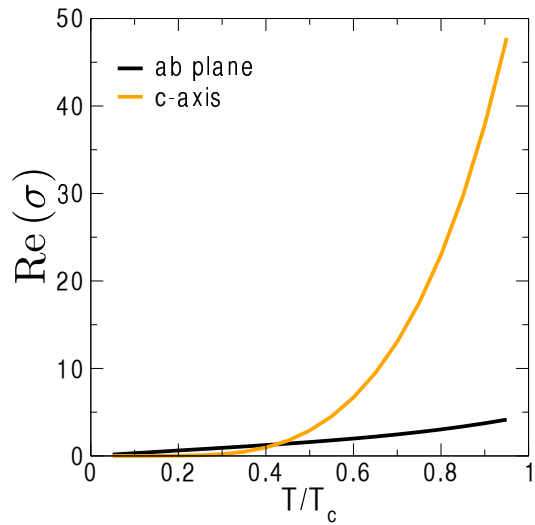}
    \caption{(Color online) Microwave conductivity (Re($\sigma_{ab}$) measured in units of $e^2/d$, and Re($\sigma_c$) measured in units of $e^2dt_\perp^2$) for a constant $\Gamma = 0.005$ as a function of temperature.  This is proportional to the number particles able to absorb low frequency radiation. The behavior in plane is linear in $T$, while the c-axis behaves roughly like $T^5$.}
  \label{fig:2}	
\end{figure}
The microwave data is qualitatively different for the in-plane and c-axis.\cite{Hosseini:1998ys}  The in-plane data contains a peak, while the c-axis does not.   The peak is attributed to the interplay between the quasiparticle lifetime and the amount of fluid which can absorb radiation.  As the system is cooled  below $T_c$ the quasiparticle scattering lifetime changes.  As the temperature decreases the normal component of the fluid vanishes linearly (Fig. \ref{fig:2}).  If the lifetime increases faster than the normal part of the fluid, this will lead to an increasing conductivity.   At low enough temperature, the lifetime will saturate to the value set by the residual scattering rate, and the conductivity will decrease as the remaining normal fluid condenses.  This behavior generally produces a peak in the microwave data.  It is thus, very surprising that the c-axis lacks a peak.  This was originally interpreted as evidence for incoherent c-axis transport\cite{Hosseini:1998ys}.  In this work we take a coherent model for c-axis transport and interpret the difference as a signature of how the superfluid density is changing.  This view is similar to the work of T. Xiang and collaborators\cite{Xiang:2000uq,Su:2006fk} where they obtained good agreement with the resistivity and microwave conductivity using a different phenomenological model.

The microwave conductivity is obtained from the low-frequency part of $\textrm{Re}[\sigma(\omega)]$.  We obtain the microwave conductivity by taking $\lim_{\omega\rightarrow0}\textrm{Re}[\sigma(\omega)]$.  In Fig. \ref{fig:2} we show the microwave conductivity for a constant scattering rate $\Gamma = 0.005$.  The solid black curve is $\textrm{Re}[\sigma(T,\omega=0)]$ for the ab-plane while orange is for the c-axis both as a function of reduced temperature, $T/T_c$.  Their temperature dependence is striking different as we would have expected based on the results presented in Fig. \ref{fig:4} for $N_N-N_S$ vs $T/T_c$.  For a first understanding of the experimental results for $\textrm{Re}[\sigma(T,\omega=0)]$, it is helpful to take guidance from the two fluid model.  Under the assumption that only the normal fluid component, $n$, is involved in the absorption, the expression for the Drude conductivity is
\begin{align}
\textrm{Re}[\sigma] = \frac{ne^2}{m\Gamma}.
\end{align}
Here $e$ is the electron charge and $m$ is the mass.  We see in Fig. \ref{fig:2} that $n$ is nearly linear in plane and $\propto T^5$ out of plane.

\begin{figure}
\centering
  \includegraphics[width=0.8\linewidth]{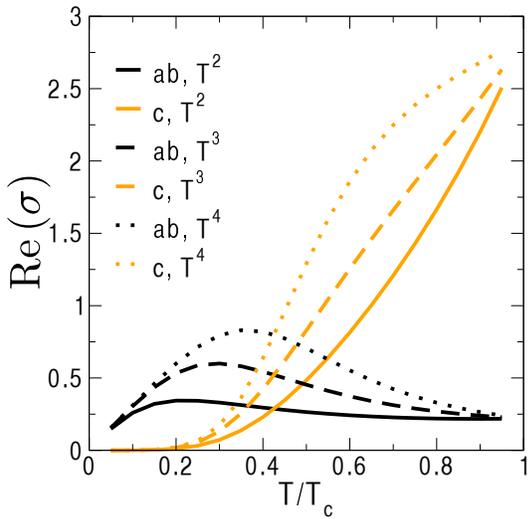}
    \caption{(Color online) The in-plane and c-axis microwave conductivity (Re($\sigma_{ab}$) measured in units of $e^2/d$, and Re($\sigma_c$) measured in units of $e^2dt_\perp^2$) for $x=0.20$ as a function of temperature for various different forms of scattering rate.}
  \label{fig:2a}	
\end{figure}

\begin{figure}
\centering
    \includegraphics[width=0.8\linewidth]{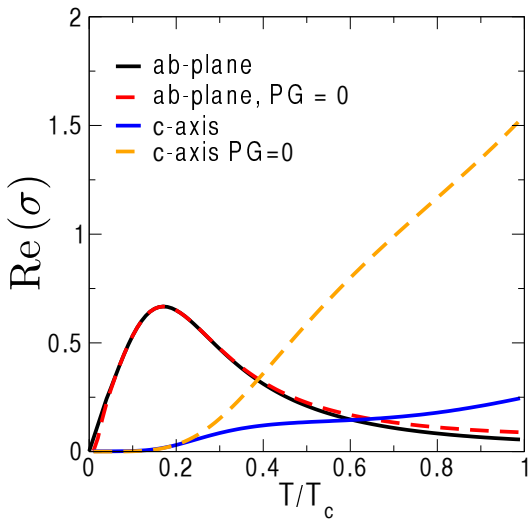}
    \caption{(Color online) The microwave conductivity (Re($\sigma_{ab}$) measured in units of $e^2/d$, and Re($\sigma_c$) measured in units of $e^2dt_\perp^2$) in the underdoped phase ($x=0.13$) as a function of temperature. We also show the calculation with $\Delta_{pg}=0$.  The pseudogap has almost no effect on the in-plane conductivity, but leads to a dramatic suppression for the c-axis.}
  \label{fig:7}	
\end{figure}

In Fig. \ref{fig:2a} we show the microwave conductivity at optimal doping for different scattering rates.  We used $\Gamma = 0.001+0.1(T/T_c)^\alpha$ for $\alpha = 2,3,4$.  Dots correspond to $T^4$ law, dashed to $T^3$ and solid to $T^2$, with black for ab and orange for c-axis.  The in-plane data always contains a peak, as we expect.  If the scattering rate becomes too strongly temperature dependent, the c-axis conductivity is no longer convex like in the experiments.  Using the Drude form for the conductivity with $\Gamma = A+BT^\alpha$, and $n \propto T^\beta$ we can show that for the conductivity to be convex that $\alpha < 1-2\beta+2\sqrt{2\beta(\beta-1)}$.  Using $\beta \approx 5$ from our constant scattering rate calculation, we see this sets an upper bound $\alpha = 3.65$.

In Fig. \ref{fig:7} we show the microwave conductivity for $x=0.13$.  To isolate the effect of the pseudogap, we redid the calculation with $\Delta_{pg} = 0$.  The solid curves are with the pseudogap and the dashed without.  The low temperature behavior is unaffected by pseudogap formation.  It does cause a dramatic suppression of the c-axis conductivity at high temperatures, but this effect is much more modest in-plane (Fig. \ref{fig:7}). There is a great similarity between the change in temperature behavior of the c-axis microwave conductivity and the specific heat.\cite{Borne:2010tg}  In both cases the low temperature part of the curve is unaffected since this region depends only on the thermal excitations in the nodal direction.  This part of the electronic structure is not appreciably changed by the pseudogap.  As the temperature is increased towards $T_c$ the specific heat is strongly suppressed below its $\Delta_{pg} =0$ value, much like the c-axis conductivity.  This is not surprising as both quantities are closely tied to the electronic density of states. The in-plane microwave conductivity is not, and is seen to behave much differently from its c-axis counterpart.  Unfortunately c-axis measurements are technically challenging and experimental data only exists for optimal doping.  It would be very interesting to look for the effect of the pseudogap in an underdoped sample.

\section{Discussion and conclusions}
\label{sec:4}
We have investigated the c-axis optical conductivity in the underdoped cuprates using the YRZ model.  We focused on properties in the superconducting phase of the underdoped cuprates at $x=0.2$ (optimal doping) and $x=0.13$ (underdoped) to highlight the essential features of the model.  For the c-axis calculations we used a coherent tunneling matrix element to describe interlayer hoping.  Our choice of matrix element is one related to the geometric alignment of atoms between adjacent layers, but any matrix element which gives little weight to states along the nodal direction should give qualitatively similar results.  We saw that the reduction in the density of states caused by the pseudogap resulted in the low frequency region of the normal state optical conductivity decreasing as temperature decreases instead of increasing as observed in-plane.  This is in agreement with the experimental findings.  This decrease continues in the superconducting state with the spectral weight redistributed to higher energies set by the pseudogap energy scale.  This redistribution could serve as an explanation for the observation of the broad peak at $400 \mathrm{cm}^{-1}$ in underdoped YBCO (YBa$_2$Cu$_3$O$_6.7$) which is not present in the optimally doped sample (YBa$_2$Cu$_3$O$_6.95$).\cite{Homes:1993fk}

We were also able to extract the behavior of the superfluid density from both a sum rule, and from the microwave conductivity.  The superfluid density behaves very differently for in-plane and out of plane, which manifested itself in the different shapes of the microwave conductivity as a function of temperature.  The in-plane microwave conductivity exhibits a peak, while the c-axis does not. A previous interpretation\cite{Hosseini:1998ys} took this observation to mean that the c-axis interplane transport was incoherent. Here we attribute this to the very different temperature law associated with the c-axis superfluid density as compared to the ab-plane.  The shape of the c-axis conductivity constrains the temperature dependence of the scattering in the superconducting state.  We also showed that the pseudogap suppresses the microwave conductivity at high temperatures. It would be interesting to see more c-axis measurements, as they display pseudogap physics more strongly than the in-plane counterparts.

\begin{acknowledgements}
This work was supported by the Natural Sciences and Engineering Research Council of Canada and the Canadian Institute for Advanced Research.
\end{acknowledgements}

\bibliography{biblio}

\end{document}